\documentstyle[preprint,prd,aps,epsfig,12pt]{revtex}
\tightenlines
\begin{document}
\draft
\preprint{
\vbox{
\halign{&##\hfil\cr
}}
}
\title{\LARGE{Semileptonic Decays of $B_c$ Meson to a $P$-Wave
Charmonium State $\chi_c$ or $h_c$}}
\author{Chao-Hsi Chang$^{a, b}$, Yu-Qi Chen$^{a,b}$,
Guo-Li Wang$^{b,c}$, Hong-Shi Zong$^{b,d}$}
\address{$^a$ CCAST (World Laboratory), P.O. Box 8730, Beijing 100080,
China\footnote{Not post mail address for the authors.}}
\address{$^b$  Institute of Theoretical Physics, Academia Sinica, P.O.
Box 2735, Beijing 100080, China}
\address{$^c$  Department of physics, FuJian Normal University, FuZhou
350007, China}
\address{$^d$  Department of physics, Nanjing University, Nanjing
210008, China}
\maketitle
\begin{center}
\begin{abstract}
The semileptonic decays of meson $B_c$ to a P-wave charmonium
state $\chi_c(^3P_J)$ or $h_c(^1P_1)$ are computed. The results
show that the decays are sizable so they are accessible in
Tevatron and in LHC, especially, with the detectors LHCB and BTeV in
the foreseeable future, and of them, the one to the $^1P_1$
charmonium state potentially offers us a novel window to see the
unconfirmed $h_c$ particle. In addition, it is pointed out that
since the two charmonium radiative decays $\chi_c(^3P_{1,2}) \to
J/\psi+\gamma$ have sizable branching ratios, the cascade decays
of the concerned decays and the charmonium radiative decays may
affect the result of the observing the $B_c$ meson through the
semileptonic decays $B_{c}{\rightarrow} {J/\psi}+{l}+\nu_{l}$
substantially.
\end{abstract}
\end{center}


\indent

The observation of the meson $B_c$ through the semi-leptonic
decays $B_{c}{\longrightarrow}{J/\psi}+{l}+\nu_{l}$ has been
reported very recently by CDF group\cite{cdf}. According to the
observation, the mass is $m_{B_c} = 6.40\pm 0.39 \pm 0.13$ GeV,
the lifetime $\tau_{B_c} = 0.46^{+0.18}_{-0.16} \pm 0.03$ ps etc.
Nevertheless, all falls in the predicted region, if the
theoretical uncertainties and experimental errors are taken into
account\cite{cdf,cch01,cchp,cchd,dec,eich}. In this letter, we
briefly report the novel results of the estimates on the
semileptonic decays of the meson $B_c$ to a P-wave chamonium
$B_{c}\to \chi_c (h_c) + {l}+\nu_{l}$ (here $\chi_c$ are the spin
triplet states $(c\bar c)[^3P_{0,1,2}]$ and $h_c$ the singlet
$(c\bar c)[^1P_1]$). Besides the decays themselves, especially, as
reported in the letter being accessible in foreseeable future, are
interesting, the cascade decays $B_c\to \chi_c(^3P_{1,2})
+l+\nu_l$ and $\chi_c(^3P_{1,2}) \to J/\psi+\gamma$ followed may
affect the result of observing the meson $B_c$ through
semileptonic decays as done in CDF substantially, it is because
the radiative decays $\chi_c(^3P_{1,2}) \to J/\psi+\gamma$ are
sizable.

In order to obtain reliable results, suitable approach to compute
the decays should be chosen. It is well known that QCD inspired
potential model works very well for non-relativistic double-heavy
systems. The non-relativistic double heavy systems ($c\bar{b}$)
and ($\bar{c} b$), except the reduce mass, are similar to the
well-studied systems ($b \bar{b}$) and ($c \bar{c}$), so it is
believed that the static properties of the systems ($c\bar{b}$)
and ($\bar{c} b$) can be predicted by potential model so well as
the systems $(b \bar{b})$ and $(c \bar{c})$. To apply the wave
functions obtained by the potential model to compute the concerned
decays for present purposes is attracting. Whereas, since
$m_{B_c}\sim 6.4$ GeV and $m_{\chi_c}\sim 3.5$ GeV, to deal with a
possible and great, even relativistic, momentum recoil effects
properly in the decays, extra special efforts are needed. As done
in computing the $B_c$ meson decays to a S-wave charmonium
($\eta_c, J/\psi$ or $\psi'$) in Ref. \cite{cchd}, the approach,
`generalized instantaneous approximation', are adopted herre. The
key points of the approach are firstly to set the potential model
on the ground of Bethe-Salpeter (B.S.) equation to describe the
binding systems instead of Shr\"odinger equation, then by means of
the Mandelstam method\cite{man}, to establish the weak current
matrix-element of the relevant process, i.e. the weak current
sandwiched by two bound-state wave functions in B.S. formulation.
Thus the current matrix element is written in a fully relativistic
formulation. The third step is to make the so-called `generalized
instantaneous approximation' on the whole relativistic matrix
element. Finally as a result of the approach, the matrix element
turns to be formulated by means of the Schr\"odinger wave
functions with proper operators sandwiched, and the Schr\"odinger
wave functions are just the ones obtained by potential model for
the systems, which are direct related to the B.S. wave functions
by the original instantaneous approximation\footnote{It is shown
by Salpeter first in early 50's.}. Note here that since the
approach starts with B.S. equation for the binding systems and the
Mandelstam formulation for the transition matrix element, the
approach has more solid ground on the quantum field theory.

Due to the restriction on the lengthy for a letter, beyond a brief
report we cannot describe the details of the estimates and on more
decay modes here, but will put them in another achieved
paper\cite{wglc}.

For the exclusive semileptonic decays $B_c\to
X_c+\ell^{+}+\nu_{\ell}$, the $T-$matrix element:
\begin{equation}
T=\frac{G_F}{\sqrt{2}}V_{ij}\overline{u}_{\nu_{\ell}}\gamma_{\mu}
(1-\gamma_5)v_{\ell}<X_c(p', \epsilon)\vert J^{\mu}\vert B_c(p)>,
\end{equation}
where $X_c$ indicates $\chi_c$ and $h_c$,
$\overline{u}_{\nu_{\ell}}$ and $v_{\ell}$ are the spinors of the
leptons, $V_{ij}$ is the Cabibbo-Kobayashi-Maskawa(CKM) matrix
element and $J^{\mu}$ is the charged weak interaction current
responsible for the decay, $p$, $p'$ are the momenta of initial
state $B_c$ and final state $X_c$. Thus we have:
\begin{equation}
\sum\vert T\vert^2=\frac{G^2_{F}}{2}\vert V_{ij}\vert^2 l^{\mu\nu}h_{\mu\nu},
\end{equation}
where $l^{\mu\nu}$, being the leptonic tensor, can be calculated
straightforwardly, and $h_{\mu\nu}$, being the hadronic tensor,
can be written as:

$$ h_{\mu\nu}=-\alpha
g_{\mu\nu}+\beta_{++}(p+p')_{\mu}(p+p')_{\nu}+
\beta_{+-}(p+p')_{\mu}(p-p')_{\nu}+\beta_{-+}(p-p')_{\mu}(p+p')_{\nu}+
$$
\begin{equation}
\beta_{--}(p-p')_{\mu}(p-p')_{\nu}+
i\gamma\epsilon_{\mu\nu\rho\sigma}(p+p')^{\rho}(p-p')^{\sigma}\;,
\end{equation}
and with the leptonic and hadronic tensors the differential decay
rate may be obtained:
\begin{center}
$$\frac{d^3\Gamma}{dxdy}=\vert
V_{ij}\vert^2\frac{G^{2}_{F}M^5}{32\pi^3} \left\{
\alpha\frac{(y-\frac{{m^2_l}}{M^2})}{M^2}
+2\beta_{++}\left[2x(1-\frac{M'^2}{M^2}+y)-
4x^2-y\right.\right.$$
$$\left.\left.+\frac{m^2_l}{4M^2}(8x+\frac{4M'^2-m^2_l}{M^2}-3y)\right]
\left.+4(\beta_{+-}+\beta_{-+})\frac{{m^2_l}}{M^2}
(2-4x+y-\frac{2M'^2-m^2_l}{M^2})\right.\right.$$
\begin{equation}\left.+
4\beta_{--}\frac{{m^2_l}}{M^2}(y-\frac{{m^2_l}}{M^2})
-\gamma\left[ y(1-\frac{M'^2}{M^2}-4x+y)+\frac{{m^2_l}}{M^2}
(1-\frac{M'^2}{M^2}+y)\right]\right\},
\end{equation}
\end{center}
where $x\equiv E_{\ell}/M$ and $y\equiv (p-p')^2/M^2$, $M, m_l$
are the masses of $B_c$ meson and the charged lepton respectively,
$M'$ is the mass of final state $X_c$. The coefficient functions
$\alpha$, $\beta_{++}$, $\gamma$ are related to the form factors
which are defined in different cases respectively:

1. In the case $X_c$ is $h_c([^1P_1])$ state,

the vector current matrix element can be written as:
\begin{equation}
<X_c(p', \epsilon)\vert V_{\mu}\vert B_c(p)>\equiv
r\epsilon^{*}_{\mu}+ s_{+}(\epsilon^{*}\cdot
p)(p+p')_{\mu}+s_{-}(\epsilon^{*}\cdot p)(p-p')_{\mu},
\end{equation}
and the axial vector current matrix element:
\begin{equation}
<X(p', \epsilon)\vert A_{\mu}\vert B_c(p)>\equiv iv
\epsilon_{\mu\nu\rho\sigma}\epsilon^{*\nu}(p+p')^{\rho}(p-p')^{\sigma}\;,
\end{equation}
where $p$ and $p'$ are the momenta of $B_c$ and $h_c$
respectively, $\epsilon$ is the polarization vector of $h_c$.

2. In the case $X_c$ is $\chi_c([^3P_0])$ state,

the axial current
\begin{equation}
<X_c(p')\vert A_{\mu}\vert B_c(p)>\equiv
u_{+}(p+p')_{\mu}+u_{-}(p-p')_{\mu}\;.
\end{equation}
The vector matrix element vanishes.

3. In the case $X_c$ is $\chi_c([^3P_1])$ state,

\begin{equation}
<X_c(p', \epsilon)\vert V_{\mu}\vert B_c(p)>\equiv
l\epsilon^{*}_{\mu}+ c_{+}(\epsilon^{*}\cdot p)(p+p')_{\mu}+
c_{-}(\epsilon^{*}\cdot p)(p-p')_{\mu}\;,
\end{equation}
\begin{equation}
<X_c(p', \epsilon)\vert A_{\mu}\vert B_c(p)>\equiv iq
\epsilon_{\mu\nu\rho\sigma}\epsilon^{*\nu}(p+p')^{\rho}(p-p')^{\sigma}\;.
\end{equation}

4. In the case $X_c$ is $\chi_c([^3P_2])$ state,

\begin{equation}
<X_c(p', \epsilon)\vert A_{\mu}\vert B_c(p)>\equiv
k\epsilon^{*}_{\mu\nu}p^{\nu}+
b_{+}(\epsilon^{*}_{\rho\sigma}p^{\rho}p^{\sigma})(p+p')_{\mu}+b_{-}
(\epsilon^{*}_{\rho\sigma}p^{\rho}p^{\sigma})(p-p')_{\mu}
\end{equation}
and
\begin{equation}
<X_c(p', \epsilon)\vert V_{\mu}\vert B_c(p)>\equiv ih
\epsilon_{\mu\nu\rho\sigma}
\epsilon^{*\nu\alpha}p_{\alpha}(p+p')^{\rho}(p-p')^{\sigma}\;.
\end{equation}

\begin{figure}\begin{center}
   \epsfig{file=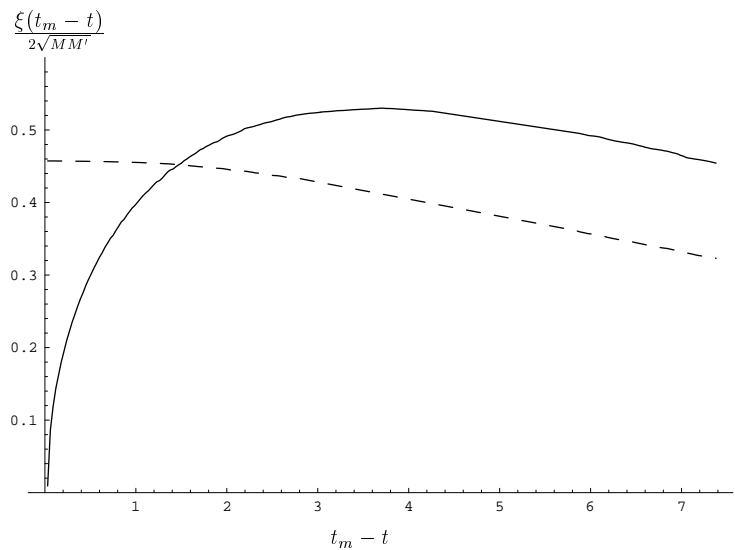, bbllx=160pt,bblly=350pt,bburx=550pt,bbury=660pt,
width=10cm,angle=0} \caption{The universal function of $\xi_1$ and
$\xi_2$: overlap-integrations of $\chi_c(h_c)$ and $B_c$ wave
functions, where the solid line is of $\xi_1$, the dashed line is
of $\xi_2$.}
\end{center}
\end{figure}

The form factors $r, s_+, s_-, v, u_+, u_-, l, c_+, c_-, k, b_+,
b_-$ and $h$ are functions of the momentum transfer $(p-p')^2$ and
can be calculated precisely with the generalized instantaneous
approximation approach\cite{cchd}. Similar to the heavy mesons
where all factors depend on only one function (Isgur-Wise
function), here all the form factors will depend on two
independent functions $\xi_1$, $\xi_2$ with certain kinematics
factors (see the example for $h_c$ below) instead, and the
functions $\xi_1$ and $\xi_2$ are just two different
overlap-integrations of the wave functions of the initial and
final bound states:
$${\epsilon}^{\lambda}(L)\cdot\epsilon_0\xi_1\equiv
\int\frac{d^3q'_{p'\perp}}{(2\pi)^3}
\psi'^{*}_{n1M_z}(q'_{p'T})\psi_{n00}(q_{pT}),$$
\begin{equation}
\epsilon^{\alpha}_{\lambda}(L)\xi_2\equiv\int\frac{d^3q'_{p'\perp}}{(2\pi)^3}
\psi'^{*}_{n1M_z}(q'_{p'T})\psi_{n00}(q_{pT})
q'^{\alpha}_{p'\perp},\label{q9}
\end{equation}
where $$\epsilon_0\equiv\frac{p-\frac{p\cdot p'}{M'^2}p'}{\sqrt
{\frac{(p\cdot p')^2}{M'^2}-M^2}}$$
describes the polarization
vector along recoil momentum $\stackrel {\rightarrow}{p'}$, and
$\epsilon^{\lambda}(L)$ is polarization vector for orbital angular
momentum.

The two functions $\xi_1$ and $\xi_2$ take into account the recoil
effects properly indeed. Note especially that $\xi_1$ cannot be
obtained by boosting the final state wave function as done in the
cases when recoil is small in atomic and nuclear processes
($\xi_1$ approaches to zero when the recoil momentum vanishes).

To see the behavior, let us present the functions $\xi_1$ and
$\xi_2$ versus $t_m-t$ in Fig.1, where $t_m=(M-M')^2$ and
$t=(P-P')^2$.

The precise formula of the form factors on the functions $\xi_1,
\xi_2$ in various cases, being lengthy for the letter, cannot be
written down here, we put them elsewhere in Ref.\cite{wglc}, but
to see the general features, only the form factors $r, s_+, s_-$
and $v$ for $h_c$ are given here:
\begin{equation}
r=\frac{(m'_1-m_2)(m_1+\omega_{10}-m_2-\omega_{20})
\xi_2}{8m'_1\omega_{10}\omega_{20}}
-\frac{(m'_1+m_2)(m_1+\omega_{10}+m_2+\omega_{20}) \xi_2(p\cdot
p')}{8M'Mm'_1\omega_{10}\omega_{20}}\; , \label{q19}
\end{equation}
\begin{eqnarray}
\nonumber
s_{+}&=&\frac{m_2[M(m_2+\omega_{20}-m_1-\omega_{10})-M'(m_1+\omega_{10}+
m_2+\omega_{20})]\xi_2}{8M'M^2\omega_{10}{\omega_{20}}^2} \\
\nonumber
&+&\frac{m_2[M(m_2+\omega_{20}-m_1-\omega_{10})-M'(m_1+\omega_{10}+
m_2+\omega_{20})]\xi_1}{8M'M\omega_{10}\omega_{20}nep} \\
&+&\frac{m_2[M(m_2\omega_{20}+\omega_{20}^2-m_1\omega_{20}-\omega_{10}^2)
-M'(m_1\omega_{20}-\omega_{10}^2+
m_2\omega_{20}+\omega_{20}^2)]\xi_2}{8M'M^2\omega_{10}^3\omega_{20}}
\\ \nonumber
&+&\frac{(m'_1+m_2)(m_1+\omega_{10}+
m_2+\omega_{20})\xi_2}{16M'Mm'_1\omega_{10}\omega_{20}}-
\frac{(M'-M)\xi_1}{8M'M^2\omega_{10}}\;, \label{q20}
\end{eqnarray}
\begin{eqnarray}
\nonumber
s_{-}&=&\frac{m_2[-M(m_2+\omega_{20}-m_1-\omega_{10})-M'(m_1+\omega_{10}+
m_2+\omega_{20})]\xi_2}{8M'M^2\omega_{10}{\omega_{20}}^2} \\
\nonumber
&+&\frac{m_2[-M(m_2+\omega_{20}-m_1-\omega_{10})-M'(m_1+\omega_{10}+
m_2+\omega_{20})]\xi_1}{8M'M\omega_{10}\omega_{20}nep} \\
&+&\frac{m_2[-M(m_2\omega_{20}+\omega_{20}^2-m_1\omega_{20}-\omega_{10}^2)
-M'(m_1\omega_{20}-\omega_{10}^2+
m_2\omega_{20}+\omega_{20}^2)]\xi_2}{8M'M^2\omega_{10}^3\omega_{20}}\\
\nonumber
&-&\frac{(m'_1+m_2)(m_1+\omega_{10}+
m_2+\omega_{20})\xi_2}{16M'Mm'_1\omega_{10}\omega_{20}}+
\frac{(M'-M)\xi_1}{8M'M^2\omega_{10}}\; , \label{q21}
\end{eqnarray}
\begin{equation}
v=-\frac{(m'_1+m_2)(m_1+m_2+\omega_{10}+\omega_{20})
\xi_2}{16M'Mm'_1\omega_{10}\omega_{20}}\; . \label{q22}
\end{equation}
Here $m_1, m_2$ are `effective masses' of $\bar b$-quark and
$c$-quark in the meson $B_c$ and $m_1', m_2$ are those of $\bar
c$-quark and $c$-quark in the $p$-wave charmonium state
respectively.

For convenience, here we have also introduced the auxiliary
parameters:

$$\omega_{20}\equiv \omega'_{2}\frac{p\cdot p'}{MM'}$$

$$ \omega_{10}\equiv \sqrt{\omega^{2}_{20}-m^2_2+m^2_1}$$

$$nep={\sqrt {\frac{(p\cdot p')^2}{M'^2}-M^2}}$$

Taking the example of $h_c$ too, here let us only present the
precise dependence of the functions $\alpha$, $\beta_{++}$ and
$\gamma$ appearing in Eq.(3) on the relevant form factors:
\begin{equation}
\alpha=r^2+4M^2\stackrel{\rightarrow}{p'}^2v^2,
\end{equation}
\begin{center}
\begin{equation}
\beta_{++}=\frac{r^2}{4M'^2}-M^2yv^2+\frac{1}{2}\left[
\frac{M^2}{M'^2}(1-y)-1\right]rs_{+}+M^2\frac{\stackrel{\rightarrow}
{p'}^2}{M'^2}s^{2}_{+},
\end{equation}
$$ \beta_{+-}=-\frac{r^2}{4M'^2}+(M^2-M'^2)v^2+\frac{1}{4}\left[
-\frac{M^2}{M'^2}(1-y)-3\right]rs_{+}$$
\begin{equation}+\frac{1}{4}\left[
\frac{M^2}{M'^2}(1-y)-1\right]rs_{-}+M^2\frac{\stackrel{\rightarrow}
{p'}^2}{M'^2}s_{+}s_{-},
\end{equation}
\begin{equation}
\beta_{-+}=\beta_{+-}
\end{equation}
\begin{equation}
\beta_{--}=\frac{r^2}{4M'^2}+\left[M^2y-2(M^2+M'^2)\right]v^2+\frac{1}{2}\left[
-\frac{M^2}{M'^2}(1-y)-3\right]rs_{-}+M^2\frac{\stackrel{\rightarrow}
{p'}^2}{M'^2}s^2_{-},
\end{equation}
\begin{equation}
\gamma=2rv.
\end{equation}
\end{center}

In the other cases for $\chi_c[^3P_{0,1,2}]$, the situation is
similar. Now with Eq.(4) and the functions $\alpha, \beta_{++},
\beta_{+-}, \beta_{-+}, \beta_{--}$ and $\gamma$ in each case for
$h_c$ and $\chi_c[^3P_{0,1,2}$, the differential decay rates may
be obtained respectively for all the semileptonic decays.

In our numerical calculations, the values of the relevant
parameters are taken from ref.\cite{data}: $V_{bc}=0.04$,
$m_c=1.846$ GeV, $m_b=5.243$ GeV, $M'=3.497$ GeV (the average
value of the mass of $\chi_c$ and $h_c$) and $M=6.213$ GeV (the
mass of $B_c$). The wave functions of $B_c$ and $\chi_c, h_c$ are
taken from potential model with suitable parameters\cite{eich}.

The integrated widths for the semileptonic decays $B_c\rightarrow
h_c(\chi_c)+e(\mu, \tau)+\nu$, as final results, are shown in
Table I.
\begin{center}
{\bf Table I. Decay widths with the unit $10^{(-15)}$ GeV}

\vspace{2mm}

\begin{tabular}{|c|c|c|c|c|}
\hline
 { } & $\Gamma(B_{c}{\rightarrow}{h_c{\small [^1P_1]}} {\ell}{ {\nu}}_{\ell})$ &
$\Gamma(B_{c}{\rightarrow}{\chi_c[^3P_0]} {\ell}{ {\nu}}_{\ell})$
& $\Gamma(B_{c}{\rightarrow}{\chi_c[^3P_1]} {\ell}{
{\nu}}_{\ell})$&$\Gamma(B_{c}{\rightarrow}{\chi_c[^3P_2]} {\ell}{
{\nu}}_{\ell})$ \\ \hline

$e(\mu)$&2.509&1.686&2.206&2.732\\ \hline

$\tau$&0.356&0.249&0.346&0.422\\ \hline
\end{tabular}
\end{center}
In the table, the first line means those of $l=e,\mu$ and the
second one means that of $l=\tau$.

To see the further feature of the decays, we present the lepton
energy spectra for $e, \mu$ only in Fig.2, where
$|\vec{p}_{\ell}|$ is the momentum of lepton.
\begin{figure}\begin{center}
   \epsfig{file=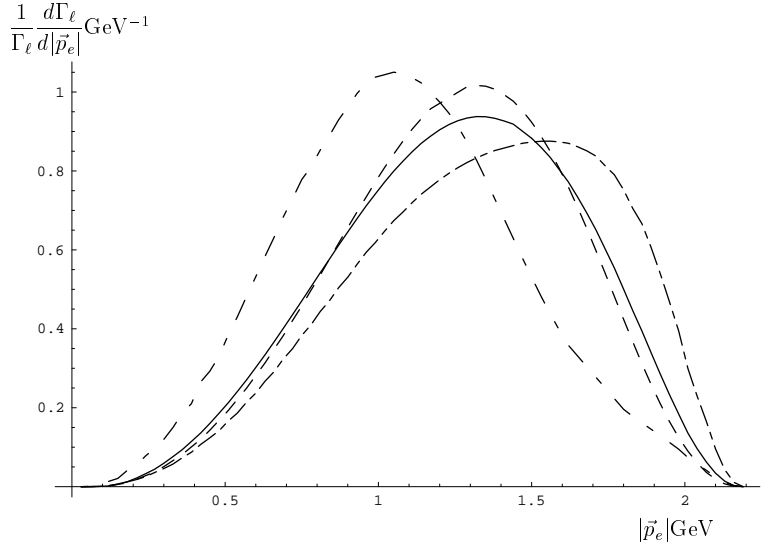, bbllx=160pt,bblly=350pt,bburx=550pt,bbury=660pt,
width=10cm,angle=0} \caption{{The lepton energy spectrum for
$B_c\rightarrow \chi_c+e(\mu)+\nu$, where the solid line is the
result of $h_c[^1P_1]$ state, the dotted-blank-dashed line is of
$\chi_c[^3P_0]$, the dashed line is of $\chi_c[^3P_1]$, the
dotted-dashed line is of $\chi_c[^3P_2]$.}}
\end{center}
\end{figure}

If comparing the results in Table 1 with the decays of $B_c$ to
$S$-wave charmonium $J/\psi$ and $\eta_c$ ($\Gamma(B_c\to
J/\psi+l+\nu)\sim 25\cdot 10^{-15}$GeV, ref.\cite{cchd,dec}) and
considering the fact about the observation of $B_c$ by CDF group,
one can conclude that the decays concerned here are accessible in
Tevatron and in LHC, especially, in the detectors LHCB and BTeV in
the foreseeable future. If considering the fact that the $P$-wave
charmonium $h_c[^1P_1]$ has not been confirmed experimentally yet,
it is quite interesting that the $B_c$ decay to the $P$-wave state
$h_c$, in fact, potentially offers us a novel window to see
it. Furthermore,
since the two charmonium radiative decays $\chi_c(^3P_{1,2}) \to
J/\psi+\gamma$ have sizable branching, the cascade decays i.e. the
decays $B_c\to \chi_c[^3P_{1,2}]+l+\nu_l$ with an accordingly
radiative one $\chi_c[^3P_{1,2}] \to J/\psi+\gamma$, being a
substantial background, may make certain affects on the result of
the observation of the $B_c$-meson through its semileptonic
decays.

\vspace{6mm}
\noindent

{\Large\bf Acknowledgement} This work was supported in part by
National Natural Science Foundation of China. The authors would
like to thank J.-P. Ma for valuable discussions. They also would
like to thank G.T. Bodwin for useful discussions.

\end{document}